\documentclass[12pt]{article}

\usepackage{scicite}
\usepackage{times}

\usepackage{xcolor}
\usepackage{amsmath}
\usepackage{bm}
\usepackage{graphicx}
\usepackage{enumitem}
\usepackage{braket}

\newenvironment{sciabstract}{%
\begin{quote} \bf}
{\end{quote}}

\title{Room temperature strain-induced Landau levels in graphene on a wafer-scale platform}

\author
{P.~Nigge$^{1,2}$, A.~C.~Qu$^{1,2}$, \'E.~Lantagne-Hurtubise$^{1,2}$, E.~M\r{a}rsell$^{1,2,3}$, S.~Link$^{4}$,\\ G.~Tom$^{1,2}$, M.~Zonno$^{1,2}$, M.~Michiardi$^{1,2,6}$, M.~Schneider$^{1,2}$,\\ S.~Zhdanovich$^{1,2}$, G.~Levy$^{1,2}$, U.~Starke$^{4}$,  C.~Guti\'errez$^{1,2}$, D.~Bonn$^{1,2}$,\\ S.~A.~Burke$^{1,2,5\ast}$, M.~Franz$^{1,2\ast}$, A.~Damascelli$^{1,2\ast}$\\
\\
\normalsize{$^{1}$Department of Physics \& Astronomy, University of British Columbia, Vancouver,}\\
\normalsize{British Columbia V6T~1Z1, Canada}\\
\normalsize{$^{2}$Quantum Matter Institute, University of British Columbia, Vancouver,}\\
\normalsize{British Columbia V6T~1Z4, Canada}\\
\normalsize{$^{3}$Division of Molecular and Condensed Matter Physics, Department of Physics and}\\
\normalsize{Astronomy, Uppsala University, P.O. Box 516, 751 20 Uppsala, Sweden}\\
\normalsize{$^{4}$Max Planck Institute for Solid State Research, 70569 Stuttgart, Germany}\\
\normalsize{$^{5}$Department of Chemistry, University of British Columbia, Vancouver,}\\
\normalsize{British Columbia V6T~1Z1, Canada}\\
\normalsize{$^{6}$Max Planck Institute for Chemical Physics of Solids, 01187 Dresden, Germany}\\
\\
\normalsize{$^\ast$To whom correspondence should be addressed; E-mail:  saburke@phas.ubc.ca,}\\
\normalsize{franz@physics.ubc.ca, damascelli@physics.ubc.ca.}
}

\date{}

\begin{document} 


\baselineskip24pt


\maketitle 


\begin{sciabstract}
Graphene is a powerful playground for studying a plethora of quantum \linebreak phenomena. One of the remarkable properties of graphene arises when it is strained in particular geometries and the electrons behave as if they were under the influence of a magnetic field. Previously, these strain-induced pseudomagnetic fields have been explored on the nano- and micrometer-scale using scanning probe and transport measurements. Heteroepitaxial strain, in contrast, is a wafer-scale engineering method. Here, we show that pseudomagnetic fields can be generated in graphene through wafer-scale epitaxial growth. Shallow triangular nanoprisms in the SiC substrate generate strain-induced uniform fields of 41 T. This enables the observation of strain-induced Landau levels at room temperature, as detected by angle-resolved photoemission spectroscopy, and confirmed by model calculations and scanning tunneling microscopy measurements. Our work demonstrates the feasibility of exploiting strain-induced quantum phases in two-dimensional Dirac materials on a wafer-scale platform, opening the field to new applications.
\end{sciabstract}


\section*{Introduction}

Graphene, a single atomic layer of carbon arranged in a honeycomb lattice, holds great promise for numerous applications due to its remarkable mechanical, optical, and electronic properties and serves as a powerful material platform for studying relativistic Dirac fermions due to its linearly dispersing bands\cite{novoselov_two-dimensional_2005,geim_rise_2007, castro_neto_electronic_2009}. Graphene was also the first material in which a member of the striking class of macroscopic quantum phenomena\cite{anderson_observation_1995,osheroff_evidence_1972,willett_observation_1987,bardeen_theory_1957} -- the quantum Hall effect (QHE)\cite{klitzing_new_1980} -- could be observed at room temperature, when subject to large magnetic fields\cite{novoselov_room-temperature_2007}. In the quantum Hall state, charge carriers are forced into cyclotron orbits with quantized radii and energies known as Landau levels (LLs), once subjected to the influence of a magnetic field. In order to observe this effect, certain conditions must be met: The magnetic field must be large enough that the resulting spacing between LLs is larger than the thermal energy $(\Delta E_{LL}>k_BT)$; the charge carrier lifetime between scattering events must be longer than the characteristic time of the cyclotron orbit $(t_\textrm{life}>1/\omega _c)$; and the magnetic field must be uniform on length scales greater than the LL orbit. This typically mandates the need for cryogenic temperatures, clean materials, and large applied magnetic fields. Dirac fermions in graphene provide a way to lift these restrictions: Under certain strain patterns, graphene's electrons behave as if they were under the influence of large magnetic fields, without applying an actual field from outside the material\cite{guinea_energy_2010,levy_strain-induced_2010,liu_tailoring_2018,rechtsman_strain-induced_2013}. These so-called pseudomagnetic fields only couple to the relativistic electrons around the Dirac point and, under the QHE conditions above, lead to the formation of flat, quantized LLs. This has been successfully observed using a range of methods\cite{levy_strain-induced_2010,liu_tailoring_2018,rechtsman_strain-induced_2013}, but was so far restricted to small regions, which severely limits its applicability.

\section*{Results}

\begin{figure}
\makebox[\textwidth]{\includegraphics[width=183mm]{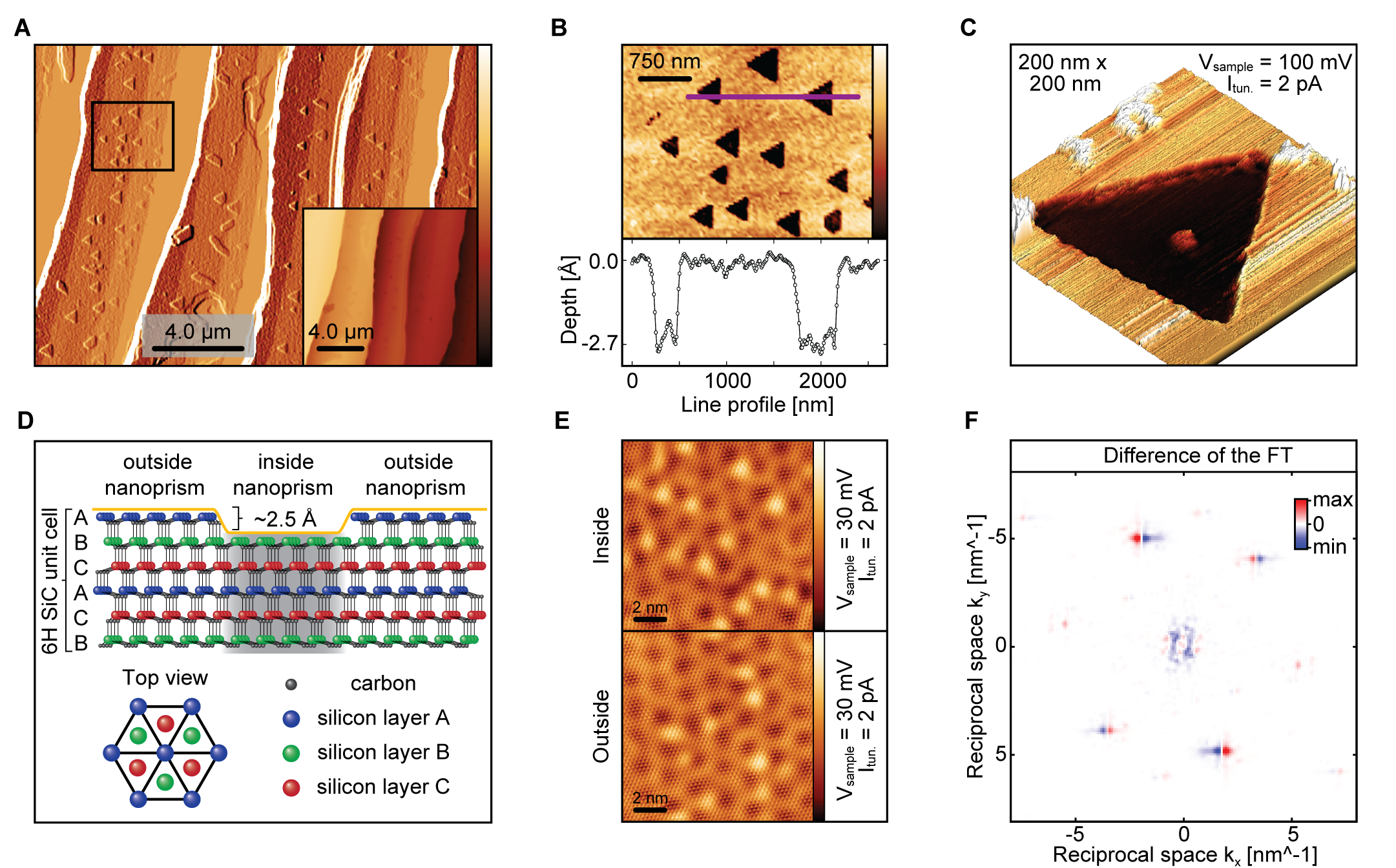}}
\caption{\textbf{Identification of strained nanoprisms.} (\textbf{A}) Horizontal derivative AFM topography image of our monolayer graphene grown on a SiC substrate. Triangular nanoprisms are dispersed on the surface. Inset: AFM topography image of the same area. Substrate terrace steps are about $10\  \textrm{nm}$ in height. (\textbf{B}) Top: Close-up AFM topography of the area indicated by the black box in (\textbf{A}). Bottom: Line cut through the AFM data marked by the purple line in the close-up. (\textbf{C}) Overview STM topography image (200 nm x 200 nm, $V_{sample}=100\ \textrm{mV}$, $I_{tun.}=2\ \textrm{pA}$) showing a single nanoprism. (\textbf{D}) Schematic structure of 6H-SiC, showing its layered ABCACB stacking order with epitaxial graphene on top (yellow). Inside the nanoprism a single layer \emph{within} the unit cell is missing, exposing the graphene to a different substrate surface termination, as illustrated in the top view. The carbon buffer layer is not shown for clarity. (\textbf{E}) Atomically resolved STM images (10\,nm x 10\,nm, $V_{sample}=30\ \textrm{mV}$, $I_{tun.}=2\ \textrm{pA}$) inside (top) and outside (bottom) of the nanoprism. (\textbf{F}) Difference map of the two Fourier transformed (FT) images in (\textbf{E}) visualizing the strain pattern inside the nanoprism.}
\label{fig:f2_afm}
\end{figure}

Here, we directly visualize the formation of flat LLs close to the Fermi energy induced by pseudomagnetic fields on wafer-scale semiconductor samples. By measuring the hallmark $\sqrt{n}$ energy spacing and momentum dependence of the ensuing pseudo-LLs with angle-resolved photoemission spectroscopy (ARPES), and with the aid of model calculations, we confirm their quantum hall nature and extract a pseudomagnetic field strength of B = 41\,T. This is made possible by the presence of a distribution of triangular nanoprisms underneath the monolayer graphene in our samples based on the well-established platform of epitaxial graphene on SiC substrates\cite{emtsev_towards_2009,riedl_quasi-free-standing_2009,bostwick_observation_2010,zhou_substrate-induced_2007}, as revealed by a combination of atomic force microscopy (AFM) and scanning tunneling microscopy (STM) measurements.\par
Our topographic images of these samples (Fig.\,\ref{fig:f2_afm}A inset) exhibit the well-known terraces and step edges of graphene grown on 6H-SiC\cite{emtsev_towards_2009}, which are due to a miscut of the wafers from the $(0001)$ direction of up to $0.1^\circ$. A population of triangular-shaped nanoscale features are identified on the terraces of our samples (Fig.\,\ref{fig:f2_afm}A), that appear similar to those reported on similar substrates\cite{bolen_graphene_2009,momeni_pakdehi_homogeneous_2019}. These nanoprisms appear during the growth process of graphene on 6H-SiC and are controllable by the Argon flow in the chamber\cite{momeni_pakdehi_homogeneous_2019}. They cover between 5\% and 10\% of the terraces and are completely covered by monolayer graphene, the latter being demonstrated by our AFM adhesion images (see Supplementary Material Figs.\,S8 and S9B). They are equilateral, have a narrow size distribution around $300\,\textrm{nm}$ side length, are oriented in the same direction, and are about $(2.7\pm 0.7)\, \textrm{\AA}$ deep (Fig.\,\ref{fig:f2_afm}B), which corresponds to a single missing SiC double layer or $\frac{1}{6}$ of the 6H-SiC unit cell. This leads to a change in the registry between the silicon atoms in the top layer of the substrate and the graphene as illustrated in Fig.\,\ref{fig:f2_afm}D. The strain created inside the nanoprisms cannot be relieved, because the nanostructures are continuously covered by monolayer graphene without additional grain boundaries, as corroborated by our STM images across the edge (see Supplementary Material Fig.\,S9A). To get a more detailed view of the strain pattern we perform additional detailed atomic resolution STM measurements. The images taken inside and outside the nanoprisms (Fig.\,\ref{fig:f2_afm}E) show the expected $(6\sqrt{3}\times6\sqrt{3})\text{R}30^\circ$ modulation with respect to SiC on top of the carbon honeycomb lattice\cite{riedl_structural_2010}. However, taking the difference of the two Fourier transformed images (Fig.\,\ref{fig:f2_afm}F) reveals a shear strain pattern inside the nanoprism, with a maximal observed strain of roughly $3^\circ$.

\begin{figure}
\makebox[\textwidth]{\includegraphics[width=184mm]{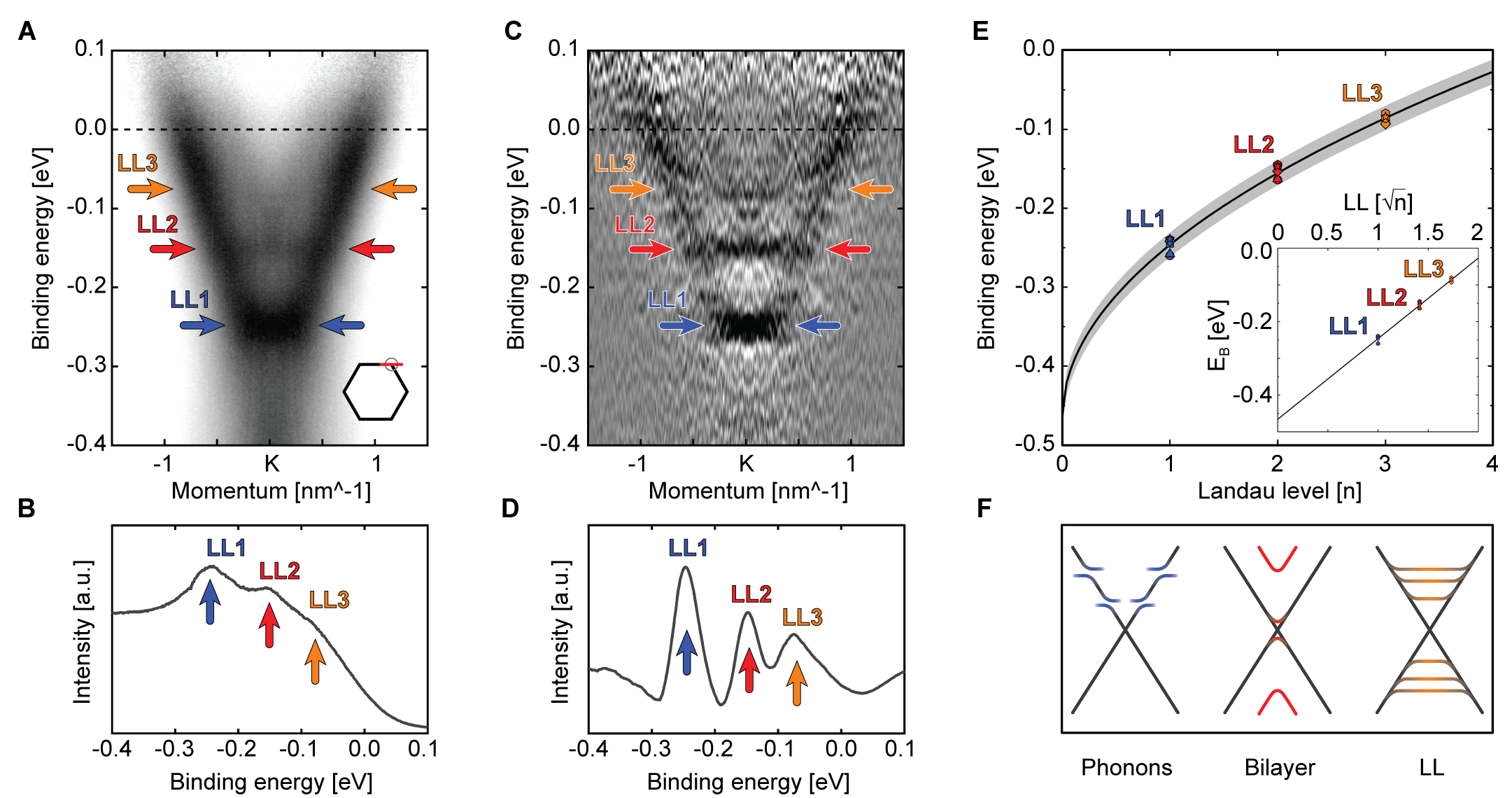}}
\caption{\textbf{Momentum-resolved visualization of Landau levels.} (\textbf{A}) ARPES cut through the Dirac cone at the K point at 300\,K. The data have been divided by the Fermi function and symmetrized to compensate for matrix element effects\cite{shirley_brillouin-zone-selection_1995}. (\textbf{B}) Cut along the energy axis integrated around the K point in (\textbf{A}). (\textbf{C}) Second derivative of the data in (\textbf{A})\cite{zhang_precise_2011}. (\textbf{D}) Inverted second derivative of the data shown in (\textbf{B}) after smoothing. (\textbf{A})--(\textbf{D}) Landau levels (LLs) are indicated by arrows. (\textbf{E}) Summary of LL data sets, with model fit according to Eqn.\,\ref{eq:landaulevels} shown in black; the 95\% confidence interval of the fit is shown in grey. Different symbols indicate different samples and temperatures: sample A (6\,K) [hexagons], sample B (6\,K) [squares], sample B 2nd data set (6\,K) [stars], sample B (300\,K) [diamonds], sample C (6\,K) [circles], and sample C 2nd data set (6\,K) [triangles]. ARPES data for the additional samples can be found in the supplementary Fig.\,S7. Inset: Same data plotted versus $\sqrt{n}$, giving the expected linear behaviour for LLs in a Dirac material. (\textbf{F}) Sketch of various mechanisms which may lead to ARPES intensity inside the cone. Neither electron-phonon coupling nor contamination from bilayer graphene can explain the experimental findings.}
\label{fig:f1_arpes}
\end{figure}

In order to confirm if the induced strain pattern indeed leads to flat Landau levels close to the Fermi energy, we perform a series of high-resolution ARPES measurements. ARPES is a momentum- and energy-resolved technique that has proven to be a powerful tool in directly studying the electronic band structures of a vast variety of quantum phases of matter, from strongly-correlated electron systems and high-T$_{c}$ superconductors\cite{damascelli_probing_2004} to topological insulators and semimetals\cite{hsieh_topological_2008,chen_experimental_2009,lv_observation_2015}. Yet no study of quantum Hall states has been performed, since ARPES is strictly incompatible with the application of magnetic fields, as essential crystal momentum information carried by the photoemitted electrons would be lost through interaction with the field. This however is different for pseudomagnetic fields, as they only interact with the Dirac electrons inside the material. We note that, while a recently developed momentum-resolved technique amenable to magnetic fields has been reported\cite{jang_full_2017}, it necessarily requires sophisticated heterostructures, physically accessible fields, and is limited to a small sector of the Brillouin zone.

Our ARPES data, which -- due to the $\sim$1\,mm spot size of the photon source -- correspond to the spatial average over unstrained and strained regions of the sample, show the expected Dirac cone as well as new flat bands that gradually merge with the linear dispersion (Figs.\,\ref{fig:f1_arpes}A and \ref{fig:f1_arpes}C). The unequal energy spacing of these newly observed bands can be extracted from cuts along the energy direction at the K point (Fig.\,\ref{fig:f1_arpes}B) and their second derivative (Fig.\,\ref{fig:f1_arpes}D). By plotting the positions of these bands (Fig.\,\ref{fig:f1_arpes}E), we observe the distinct $\sqrt{n}$ energy spacing which is a hallmark of LLs for graphene's massless Dirac charge carriers\cite{geim_rise_2007}, where $n$ is the integer LL index. The spectrum of LLs in graphene is given by\cite{castro_neto_electronic_2009}
\begin{equation}
    E_n=\text{sgn}(n)\sqrt{2v_F^2\hbar eB\cdot |n|}+E_{DP}
    \label{eq:landaulevels}
\end{equation}  
where $v_F$ is the velocity of the electrons at the Fermi level, $\hbar$ the reduced Planck constant, $e$ the electron charge, $B$ the magnitude of the (pseudo-)magnetic field, and $E_{DP}$ the binding energy of the Dirac point. Using the ARPES dispersion map in Fig.\,\ref{fig:f1_arpes}A, the Fermi velocity is determined to be $v_F=(9.50\pm0.08)\times10^5\ \textrm{ms}^{-1}$ (see Supplementary Material Fig.\,S1). Fitting our experimental data to Eqn.\,\ref{eq:landaulevels} as done in Fig.\,\ref{fig:f1_arpes}E, we extract the magnitude of the pseudomagnetic field, which yields $B=(41\pm 2)\ \textrm{T}$. Remarkably, this pseudomagnetic field value is consistent between several samples from cryogenic temperatures (6\,K) up to room temperature. The model fit also consistently pinpoints the binding energy of the Dirac point to $E_{DP}=(460\pm 10)\ \textrm{meV}$ relative to the Fermi level, which agrees well with previous reports on this sample system\cite{bostwick_quasiparticle_2007,emtsev_towards_2009} and is attributed to charge transfer from the SiC substrate to the graphene layer. Additionally, the LLs are only resolved in the upper part of the Dirac cone, closer to the Fermi level\cite{levy_strain-induced_2010, song_high-resolution_2010}. We attribute this effect to the increased scattering phase space as one moves away from the Fermi level, which manifests itself in our ARPES data by an increased line width of the bands (Supplementary Material Fig.\,S1).

As for other alternative explanations of the data, we note that while previous ARPES studies of graphene on SiC have shown a rich variety of features\cite{ludbrook_evidence_2015,ohta_controlling_2006}, the signature $\sqrt{n}$ spacing of the levels (Fig.\,\ref{fig:f1_arpes}E and inset) allows us to unambiguously distinguish the observed effect from other possibilities (Fig.\,\ref{fig:f1_arpes}F). For example, if spectral weight inside the Dirac cone arose from coupling of electrons to phonons\cite{ludbrook_evidence_2015}, it would be limited to characteristic vibrational energies. Similarly, contributions from bilayer and higher order graphene layers, which can appear in small quantities near step edges of the substrate during the growth process\cite{ohta_controlling_2006} (see also AFM adhesion image in the Supplementary Material Fig.\,S9B), would lead to a manifold of bands, but would not reproduce the observed band structure\cite{ohta_interlayer_2007,marchenko_extremely_2018}. Furthermore, previously reported plasmaronic interactions in samples with higher electronic doping \cite{bostwick_observation_2010} can also be excluded. They lead to renormalizations of electronic bands around the Dirac point, but show a distinctly different spectrum than what is observed in our experiments. Finally, the effects of different defect geometries in graphene and their influence on the Dirac cone dispersion have recently been discussed \cite{kot_band_2018}, but do not lead to flat bands around the Dirac point.

\begin{figure}
\makebox[\textwidth]{\includegraphics[width=183mm]{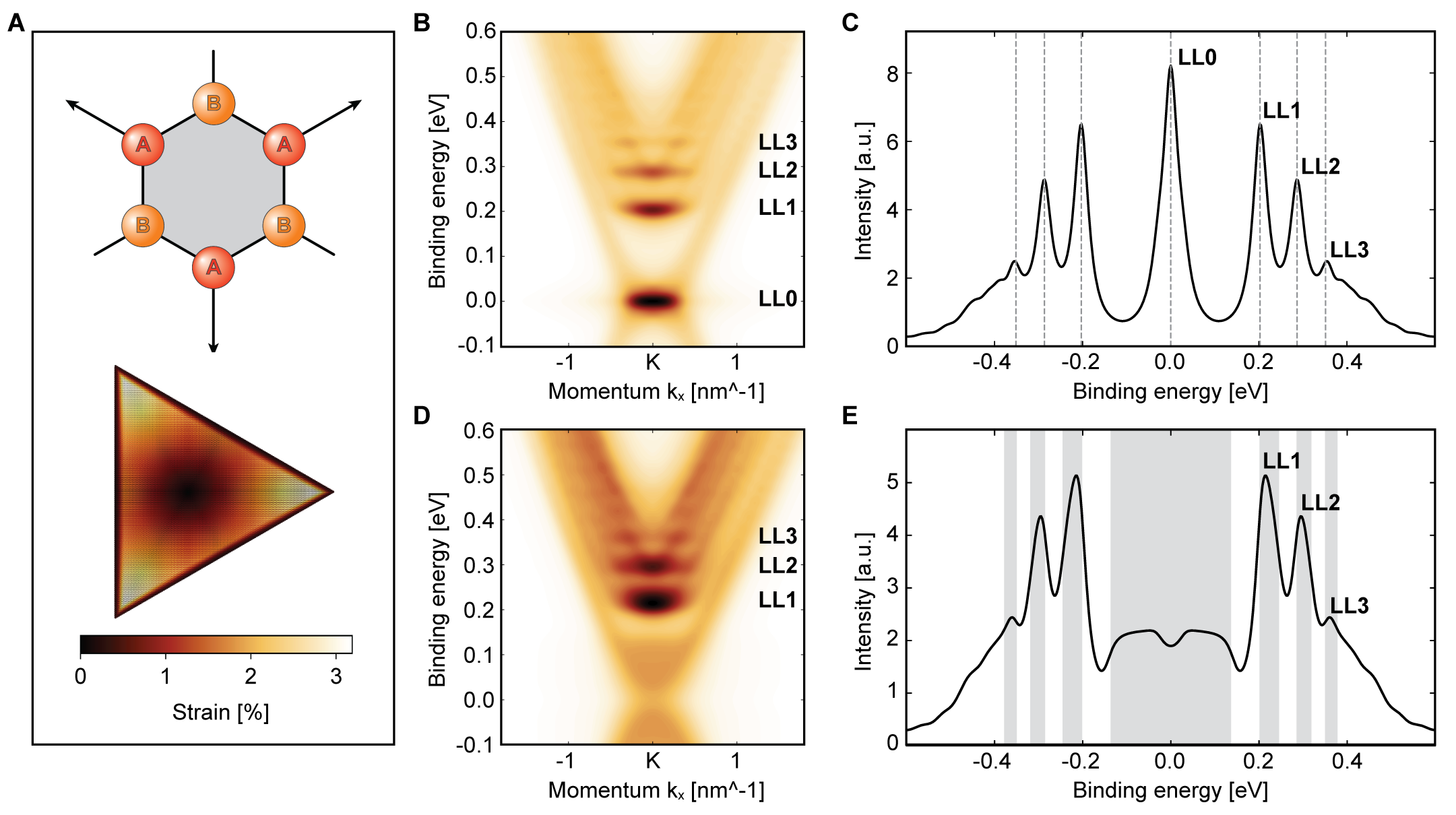}}
\caption{\textbf{Model calculation of strain-induced Landau levels.} (\textbf{A}) Top: Honeycomb lattice, with the two sublattices A (red) and B (yellow). The black arrows indicate the symmetry of the strain pattern. Bottom: Triangular flake with strain-induced pseudomagnetic field $B = 41$\,T. The colour scale indicates the relative bond stretching. (\textbf{B}) Spectral function for the gapless case with Semenoff mass $M = 0\ \textrm{meV}$. (\textbf{C}) Energy cut through the Dirac point (K) of the spectral function in (\textbf{B}). The dashed grey lines indicate the position of the Landau levels (LL) predicted by Eqn.\,\ref{eq:landaulevels}. (\textbf{D}) Spectral function averaged over a uniform distribution of Semenoff masses $M \in [-135, 135]$\,meV. (\textbf{E}) Energy cut through the Dirac point (K) of the spectral function in (\textbf{D}). The shaded grey area indicates the broadening of the Landau levels predicted by Eqns.\,\ref{eq:landaulevels} and \ref{eq:landaulevels_mass}.}
\label{fig:f3_theory}
\end{figure}

To gain deeper insights on the origin of the observed LLs, we model a region of graphene experiencing a uniform strain-induced pseudomagnetic field. We use the simplest such strain pattern, worked out by Guinea et al.\cite{guinea_energy_2010}, which exhibits the triangular symmetry of the underlying honeycomb lattice. Using a tight-binding approach, we directly simulate a finite-size strained region with open boundary conditions and armchair edges (further details in Methods section). We find that the observed LL spectra can be well reproduced by a triangular flake of side length $L = 56$\,nm (Fig.\,\ref{fig:f3_theory}A), subject to a uniform pseudomagnetic field $B = 41$\,T over the entire flake (Fig.\,\ref{fig:f3_theory}A). The maximal strain (or relative bond stretching) reaches around $3\%$, which is in good agreement with our STM measurements. The ARPES data can be simulated by calculating the energy and momentum-resolved spectral function $A(\bm{k},\omega)$ of this triangular flake, here shown in Figs.\,\ref{fig:f3_theory}B and \ref{fig:f3_theory}C. Our simulation clearly reproduces the main features of the ARPES data, namely levels that: (i) follow $\sqrt{n}$ spacing in energy; (ii) are flat \emph{inside} the Dirac cone and merge with the linearly dispersing bands; (iii) become less clearly resolved with increasing index $n$.

Features (ii) and (iii) can be understood by comparing the characteristic size of a Landau orbit $\propto\sqrt{n}\,l_B$ (with the magnetic length $l_B=\sqrt{\frac{\hbar}{eB}}$) to the length scale $\lambda$ on which the pseudomagnetic field is uniform. For LLs to exist, an electron on a given Landau orbit must experience a uniform pseudomagnetic field\cite{settnes_pseudomagnetic_2016}, leading to the condition $\sqrt{n}\,l_B \ll \lambda$. Hence, for large fields $B$ or large $\lambda$, flat bands are expected across the entire Brillouin zone, whereas Dirac cones are recovered in the opposite limit (see Supplementary Material Fig.\,S2). The bands observed in the ARPES data can thus be understood as LLs, where the orbit size is only somewhat smaller than $\lambda$: by comparing the experimental data and the model calculation, we estimate $l_B \sim 4$~nm and $\lambda \sim 30$\,nm (see Methods section). Furthermore, since the size of Landau orbits grows as $ \sim \sqrt{|n|}$, eventually it becomes comparable to $\lambda$, explaining why levels with higher index $n$ are less clearly resolved.

However, our simple model (Figs.\,\ref{fig:f3_theory}B and \ref{fig:f3_theory}C) consistently exhibits a sharp zeroth Landau level (LL0), which is absent in the ARPES data. This discrepancy is surprising, since LL0 is known to be stable against inhomogeneities of the magnetic field as well as against disorder, as long as the latter preserves the chiral symmetry of graphene\cite{aharonov_ground_1979}. Below, we provide a possible mechanism that broadens LL0 without substantially affecting the higher LLs. It has been argued that graphene grown on SiC is subject to a sublattice-symmetry-breaking potential arising from the interaction with the substrate\cite{zhou_substrate-induced_2007}. The minimal theoretical model describing this effect, which acts as a staggered potential between sublattices A and B, is the so-called Semenoff mass $M$\cite{semenoff_condensed-matter_1984}. This mass term opens a gap at the Dirac point and shifts the LL spectrum for $n\neq0$ to (for a more detailed discussion, and the particular case of $n=0$, see Supplementary Material)\cite{hunt_massive_2013}:
\begin{equation}
    E_n = \text{sgn}(n) \sqrt{ 2 v_F^2  \hbar e B\cdot|n| + M^2}+E_{DP}.
    \label{eq:landaulevels_mass}
\end{equation}
However, a \emph{uniform} mass term $M$ cannot explain the ARPES data. Indeed, a fit of the observed LL spectrum to Eqn.\,\ref{eq:landaulevels_mass} returns $M=(150\pm 5)\ \textrm{meV}$, but places the Dirac point at an unrealistic binding energy of $E_{DP}=390\ \textrm{meV}$ (see Supplementary Material Fig.\,S3 and Fig.\,S6). Therefore, we postulate that the mass term $M$ varies on a length scale much greater than the magnetic length $l_B \sim$\,4\,nm, but smaller than the ARPES spot size ($\sim$\,1\,mm). In that situation, our ARPES measurements would simply average over spectral functions described by different mass terms.  This is shown in Figs.\,\ref{fig:f3_theory}D and \ref{fig:f3_theory}E for a uniform distribution in the interval $M \in [-135, 135]$\,meV. As evident from Eqn.\,\ref{eq:landaulevels_mass}, the distribution of mass terms affects LL0 most, while merely contributing an additional broadening to the higher levels. Note that, as observed experimentally, the variation of the mass term is not limited to the strained areas, but instead is a property of the whole sample; as a result, ARPES always picks up a spatial average of strained areas with LLs and unstrained areas with the usual Dirac cone dispersion, both having the same distribution of mass terms and corresponding Dirac point gaps.  This phenomenological model is in good agreement with the experimental data and may renew interest in the variation of the mass term in this sample system\cite{zhou_substrate-induced_2007}.

\section*{Discussion}

This study provides the first demonstration of the room temperature strain-induced quantum Hall effect in graphene on a wafer-scale platform, as well as the first direct momentum-space visualization of graphene electrons in the strain-induced quantum Hall phase by ARPES, whereby the linear Dirac dispersion collapses into a ladder of quantized LLs. This opens a path for future momentum-resolved studies of strain-induced, room temperature-stable topological phases in a range of materials including Dirac and Weyl semimetals\cite{pikulin_chiral_2016,cortijo_elastic_2015,liu_quantum_2017}, monolayer transition metal dichalcogenides\cite{rostami_theory_2015}, and even nodal superconductors\cite{massarelli_pseudo-landau_2017,nica_landau_2018}, all under large, potentially controllable pseudomagnetic fields. Importantly, these systems will feature time reversal invariant ground states -- otherwise impossible with a true magnetic field -- and may act as future building blocks for pseudo spin- or valley-tronic based technologies\cite{low_strain-induced_2010}. In light of the recently discovered unconventional superconductivity in 'magic angle' bilayer graphene\cite{cao_correlated_2018,cao_unconventional_2018}, strain-induced pseudomagnetic fields likewise raise the possibility of engineering exotic variants of correlated states including superconductivity in LLs\cite{uchoa_superconducting_2013} and fractional topological phases\cite{ghaemi_fractional_2012}. Our results lay the foundations for bottom-up strain-engineering of novel quantum phases at room temperature and on a technologically relevant wafer-scale platform.

\section*{Materials and Methods}

\subsection*{Sample growth and characterization}

Graphene samples with a carbon buffer layer were epitaxially grown on commercial 6H-SiC substrates. The substrates were hydrogen-etched prior to the growth under argon atmosphere. Details are described by \emph{S. Forti and U. Starke}\cite{forti_epitaxial_2014}. AFM characterization measurements were taken at the Max Planck Institute in Stuttgart. Adhesion images correspond to the force necessary to retract the tip from the sample. Adhesion is sensitive to the graphene coverage on the sample and can thus distinguish between zero layer, monolayer and bilayer graphene with sensitivity to grain boundaries.

\subsection*{ARPES measurements}

Experiments were performed at UBC in a ultra-high vacuum chamber equipped with a SPECS Phoibos 150 analyzer with $\Delta E=6\,\textrm{meV}$ and $\Delta k=0.01\,\textrm{\AA}$ optimum energy and momentum resolutions, respectively, at a base pressure of better than $p=7\times10^{-11}\,\textrm{Torr}$. Photons with an energy of $21.2$\,eV were provided by a SPECS UVS300 monochromatized gas discharge lamp. Our homebuilt six-axis cryogenic manipulator allows for measurements between 300\,K and 3.5\,K. Additional data sets were taken at UBC with a second ARPES setup equipped with a Scienta R4000 analyzer and a Scienta VUV5000 UV source with $\Delta E=1.5\,\textrm{meV}$ and $\Delta k=0.01\,\textrm{\AA}^{-1}$ optimum energy and momentum resolutions, respectively, for $21.2\ \textrm{eV}$ photons. The samples were annealed at 600$^\circ$C for about 2\,h at $p=1\times10^{-9}\,\textrm{Torr}$ and then at 500$^\circ$C for about 10\,h at $p=5\times10^{-10}\,\textrm{Torr}$ immediately before the ARPES measurements.

\subsection*{STM measurements}

Experiments were performed at UBC under ultra-high vacuum conditions ($<5\times10^{-12}$\,mbar) using a low-temperature scanning tunnelling microscope (Scienta Omicron) at liquid helium temperatures ($\sim$\,4.2\,K). All images were acquired in constant-current mode using a cut\linebreak platinum-iridium tip, which was conditioned by voltage pulsing and gentle indentation into a Ag(111) crystal. The samples were annealed at 550$^\circ$C overnight with a final pressure of $p=3\times10^{-10}\,\textrm{mbar}$ \emph{in situ} prior to the STM measurements.

\subsection*{Model calculation}

We considered a minimal tight-binding model on the honeycomb lattice with nearest-neighbour hoppings and a sublattice-symmetry breaking Semenoff\cite{semenoff_condensed-matter_1984} mass term $M$:
\begin{equation}
H = -t \sum_{<\bm{r}, \bm{r'}>} \left( c_A^\dagger(\bm{r)} c_B(\bm{r'}) + \text{H.c.} \right) + M \left( \sum_{\bm{r}} c_A^\dagger(\bm{r}) c_A(\bm{r}) - \sum_{\bm{r'}}  c_B^\dagger(\bm{r'}) c_B(\bm{r'}) \right)
\label{eq:tight_binding}
\end{equation}
where $c^\dagger_A(\bm{r})$ ($c^\dagger_B(\bm{r'})$) creates an electron in the $p_z$ orbital at lattice site $\bm{r}$ ($\bm{r'}$) on the sublattice A (B) of the honeycomb lattice, $t = 2.7\ \text{eV}$ and the nearest-neighbour distance is $a_0 =0.142$\,nm. We neglected the electron spin, and thus considered effectively spinless fermions.

We constructed a flake in the shape of an equilateral triangle of side length $L \sim 56$\,nm. The use of armchair edges avoids the zero-energy edge modes appearing for zigzag edges\cite{castro_neto_electronic_2009}. We applied the simplest strain pattern respecting the triangular symmetry of the problem at hand, namely, the pattern introduced by \textit{Guinea et al.}\cite{guinea_energy_2010} which gives rise to a uniform (out-of-plane) pseudomagnetic field
\begin{equation}
    \bm{B} = 4 u_0 \frac{\hbar \beta}{e a_0} \hat{\mathbf{z}}
\end{equation}
where $\beta \approx 3.37$ in graphene\cite{settnes_pseudomagnetic_2016}, and the corresponding displacement field is given by
\begin{equation}
    \mathbf{u}(r,\theta) = 
    \begin{pmatrix}
    u_r \\
    u_\theta
    \end{pmatrix} 
    =
    \begin{pmatrix}
    u_0 r^2 \sin(3\theta) \\
    u_0 r^2 \cos(3\theta)
    \end{pmatrix}.
\end{equation}
The hopping parameter renormalization induced by this displacement field is calculated using the simple prescription:
\begin{align}
t \rightarrow t_{ij} = t \exp \left[ -\frac{\beta}{a_0^2} \left( \epsilon_{xx} x_{ij}^2 + \epsilon_{yy} y_{ij}^2 + 2 \epsilon_{xy} x_{ij} y_{ij} \right) \right]
\label{eq_hoppings_strain}
\end{align}
where $(x_{ij}, y_{ij}) \equiv \bm{r}_i - \bm{r}_j$ is the vector joining the original (unstrained) sites $i$ and $j$, and
\begin{equation}
    \epsilon_{ij} = \frac{1}{2} \left[ \partial_j u_i + \partial_i u_j \right]
\end{equation}
is the strain tensor corresponding to the (in-plane) displacement field $\mathbf{u}$. Outside the strained region (which we take as a triangle of slightly smaller length $L_S \sim 48$\,nm), we allowed the strain tensor to relax: $\bm{\epsilon} \rightarrow e^{-\frac{r^2}{2\sigma^2}} \bm{\epsilon}$, where $r$ is the perpendicular distance to the boundary of the strained region, and $\sigma \sim 1$\,nm. We defined the length scale of the homogeneous magnetic field $\bm{B}$ to be the diameter of the largest inscribed circle in the triangle of side $L_S$: $\lambda \equiv L_S/\sqrt{3} \sim 28$\,nm. We stress here that our simulated flakes are much smaller than the experimentally observed triangular features of size $\sim 300$\,nm. The fact that we nevertheless reproduce the experimental features underlines how the number of observable LLs is limited by the length scale of the homogeneous pseudomagnetic field $\lambda$, rather than by the size $L$ of the nanoprisms themselves. This length scale could be caused by the more complicated strain pattern present in the nanoprisms or be induced by disorder.

We then diagonalized the Hamiltonian (Eqn.\,\ref{eq:tight_binding}) with hopping parameters given by Eqn.\,\ref{eq_hoppings_strain} to obtain the full set of eigenstates $\ket{n}$ with energies $E_n$, and computed the momentum-resolved, retarded Green's function using the Lehman representation
\begin{align}
    G_\alpha^R(\bm{k},\omega) &= \sum_n \frac{ | \bra n c^\dagger_\alpha({\bm{k}}) \ket{0} |^2}{\omega - (E_n - E_0) -i \eta}
\end{align}
where $\alpha = A,B$ is a sublattice (band) index, and $\eta \sim 20$\,meV is a small broadening parameter comparable to the experimental resolution. 
%
%
%
%
%
%
We then compute the one-particle spectral function,
\begin{align}
A(\bm{k}, \omega) = -\frac{1}{\pi} \sum_\alpha \text{Im}\left[ G_\alpha^R(\bm{k}, \omega)\right]    
\end{align}
which is proportional to the intensity measured in ARPES (modulo the Fermi-Dirac distribution and dipole matrix elements). We note that using a finite system introduces two main effects in the momentum-resolved spectral function: the appearance of a small finite-size gap at the Dirac points (in the absence of a magnetic field) and a momentum broadening of the bands.


\bibliography{Graphene_LIB}
\bibliographystyle{ScienceAdvancesCite}

\nocite{emtsev_interaction_2008}
\nocite{virojanadara_homogeneous_2008}

\section*{Acknowledgments}
We gratefully acknowledge C.~Ast, G.~A.~Sawatzky, J.~Smet, J.~F.~Young, Z.~Ye, F.~Boschini, J.~Day, J.~Kim, J.~Geurs, and C.~Li for fruitful discussions. We thank D.~Wong for technical assistance and E.~Razzoli for support with data processing and analysis. This research was undertaken thanks in part to funding from the Max Planck-UBC-UTokyo Centre for Quantum Materials and the Canada First Research Excellence Fund, Quantum Materials and Future Technologies Program. The work at UBC was supported by the Killam, Alfred P. Sloan, and Natural Sciences and Engineering Research Council of Canada's (NSERC's) Steacie Memorial Fellowships (A.D.), the Alexander von Humboldt Fellowship (A.D.), the Canada Research Chairs Program (A.D. and S.A.B.), NSERC, Canada Foundation for Innovation (CFI), British Columbia Knowledge Development Fund (BCKDF), and CIFAR Quantum Materials Program. P.N. thanks the UBC 4YF scholarship and the UBC SBQMI QuEST scholarship for financial support. A.C.Q. and \'{E}.L.-H. are supported by the NSERC Alexander Graham Bell Canada Graduate Scholarships - Doctoral (CGS-D) Program. E.M. acknowledges funding from the Swedish Research Council (VR) 2016-06719.

\section*{Competing Interests}
The authors declare that they have no competing interests.

\section*{Data Availability}
All data needed to evaluate the conclusions in the paper are present in the paper and/or the Supplementary Materials. Additional data related to this paper may be requested from the authors.

\section*{Author contributions}
P.N. and A.C.Q. performed the ARPES experiments and analyzed the ARPES data. P.N., A.C.Q., E.M., and G.T. performed the STM experiments and analyzed the STM data. \'{E}.L.-H. and M.F. provided the theoretical modeling, with input from C.G. S.L. and U.S. grew the samples and performed the AFM experiment. P.N., A.C.Q., M.Z., M.M., M.S., S.Z. and G.L. provided technical support and maintenance for the ARPES setup. A.D., M.F., S.A.B., D.B., and C.G. supervised the project. P.N., A.C.Q., \'{E}.L.-H., and C.G. wrote the manuscript with input from all authors. A.D. was responsible for overall project direction, planning, and management.

\newpage

\baselineskip24pt

\setcounter{figure}
    {0}
\renewcommand{\thefigure}{S\arabic{figure}}

\section*{Supplementary materials}
Supplementary Text\\
Figs. S1 to S9\\
References \textit{(51-52)}

\vspace{10mm}
 
\begin{center}
 
{\Large \textbf{Room temperature strain-induced Landau levels in graphene on a wafer-scale platform}}

\vspace{10mm}
 
P.~Nigge$^{1,2}$, A.~C.~Qu$^{1,2}$, \'E.~Lantagne-Hurtubise$^{1,2}$, E.~M\r{a}rsell$^{1,2,3}$, S.~Link$^{4}$,\\ G.~Tom$^{1,2}$, M.~Zonno$^{1,2}$, M.~Michiardi$^{1,2,6}$, M.~Schneider$^{1,2}$,\\ S.~Zhdanovich$^{1,2}$, G.~Levy$^{1,2}$, U.~Starke$^{4}$,  C.~Guti\'errez$^{1,2}$, D.~Bonn$^{1,2}$,\\ S.~A.~Burke$^{1,2,5\ast}$, M.~Franz$^{1,2\ast}$, A.~Damascelli$^{1,2\ast}$\\

\vspace{10mm}

\normalsize{$^{1}$Department of Physics \& Astronomy, University of British Columbia, Vancouver,}\\
\normalsize{British Columbia V6T~1Z1, Canada}\\
\vspace{1.5mm}
\normalsize{$^{2}$Quantum Matter Institute, University of British Columbia, Vancouver,}\\
\normalsize{British Columbia V6T~1Z4, Canada}\\
\vspace{1.5mm}
\normalsize{$^{3}$Division of Molecular and Condensed Matter Physics, Department of Physics and}\\
\normalsize{Astronomy, Uppsala University, P.O. Box 516, 751 20 Uppsala, Sweden}\\
\vspace{1.5mm}
\normalsize{$^{4}$Max Planck Institute for Solid State Research, 70569 Stuttgart, Germany}\\
\vspace{1.5mm}
\normalsize{$^{5}$Department of Chemistry, University of British Columbia, Vancouver,}\\
\normalsize{British Columbia V6T~1Z1, Canada}\\
\vspace{1.5mm}
\normalsize{$^{6}$Max Planck Institute for Chemical Physics of Solids, 01187 Dresden, Germany}\\

\vspace{10mm}

\normalsize{$^\ast$To whom correspondence should be addressed; E-mail:  saburke@phas.ubc.ca,}\\
\normalsize{franz@physics.ubc.ca, damascelli@physics.ubc.ca.}

\end{center}

\newpage

\begin{enumerate}[label=\textbf{\arabic*.},wide, labelwidth=!, labelindent=0pt]

    \item \ \textbf{Extraction of Fermi velocity and quasiparticle lifetime from ARPES data}

     \begin{figure}[b!]
\makebox[\textwidth]{\includegraphics[width=183mm]{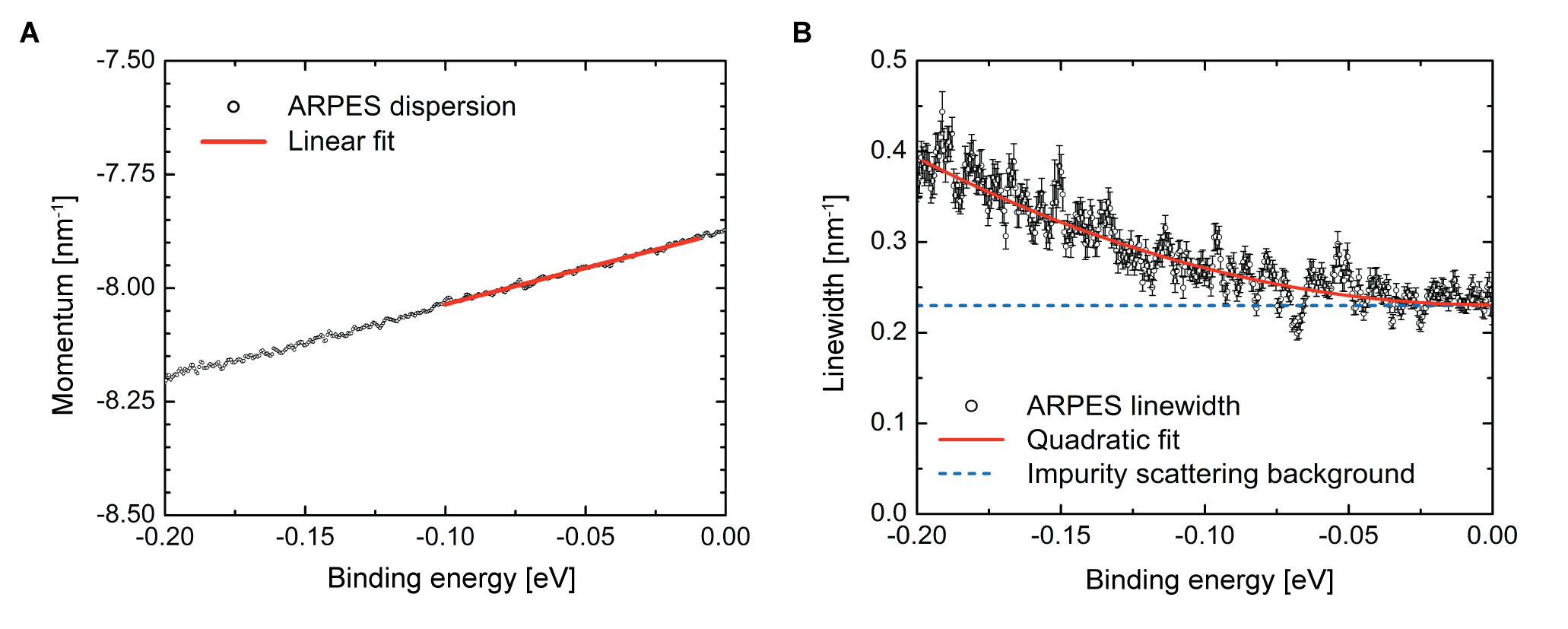}}
\caption{\textbf{Fermi velocity and quasiparticle lifetime from ARPES.} (\textbf{A}) The linear dispersion of graphene (black circles) is fitted linearly (red line) to extract the Fermi velocity. (\textbf{B}) The extracted binding energy dependent linewidth (black circles) is fitted quadratically (red line) to illustrate  the \emph{decreasing} carrier lifetime at higher binding energies. The blue dashed line indicates a constant offset due to impurity scattering.}
\label{fig:s1_fermi_velocity}
    \end{figure}

The Fermi velocity and binding energy dependence of the carriers can be directly extracted from the ARPES data. The momentum distribution curves (MDCs) at each binding energy are fitted using a Lorentzian with a constant background. Firstly, the dispersion of the band can then be fitted linearly to determine the Fermi velocity to $v_F=(9.50\pm0.08)\times10^5\ \textrm{ms}^{-1}$ (see Fig.\,\ref{fig:s1_fermi_velocity}A). Secondly, the width of the Lorentzians as a function of binding energy can be fitted quadratically with a constant offset. The linewidth is inversely proportional to the quasiparticle lifetime, thus showing how the latter \emph{decreases} as one goes away from the Fermi level (see Fig.\,\ref{fig:s1_fermi_velocity}B). This is a manifestation of a simple Fermi liquid model. Electrons at the Fermi level have a certain lifetime between scattering events dictated by the concentration of impurities and defects. As one goes to higher binding energies, the phase space for electron-electron scattering increases $\propto E_b^2$ and the lifetime decreases. We propose this as the reason why, experimentally, our LLs are only clearly resolved in the upper part of the cone closer to the Fermi level. When the scattering rate at some binding energy exceeds a critical value above which coherent circular orbits cannot be established, the LL quantization in the ARPES measurement disappears. We note that such asymmetric behaviour has been reported before in scanning probe measurements, and was attributed to a shorter vertical extension of wave functions at lower energies\,\cite{levy_strain-induced_2010} and to a reduced quasiparticle lifetime away from the Fermi level as well\cite{song_high-resolution_2010}.

     \item \ \textbf{Evolution of LLs with magnetic field strength}

\begin{figure}[b!]
\makebox[\textwidth]{\includegraphics[width=183mm]{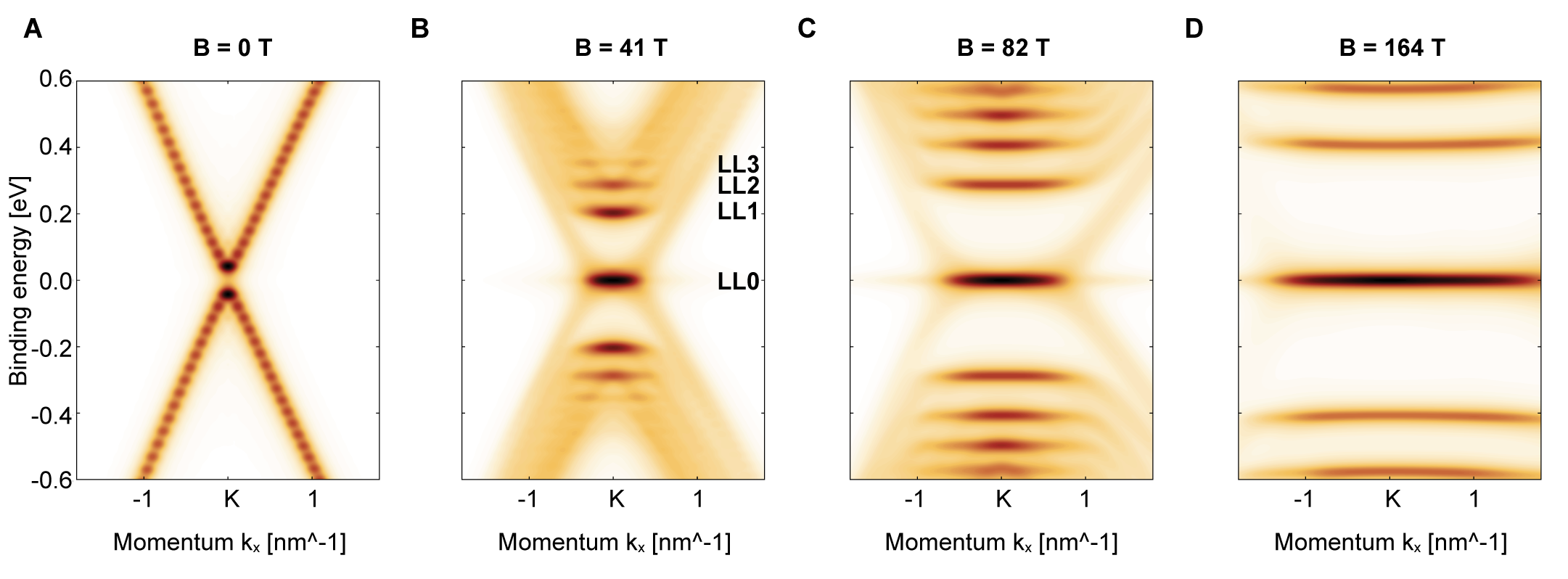}}
\caption{\textbf{Evolution of LLs for increasing uniform pseudomagnetic fields}. Calculated spectral function in our triangular flake for fields $B=0$, $41$, $82$ and $164$\,T (from left to right).}
\label{fig:s7_field_evolution}
    \end{figure}

In Fig.\,\ref{fig:s7_field_evolution}, we present the spectral function obtained for $M=0$ and increasing pseudomagnetic fields $B = 0$, $41$, $82$ and $164$\,T to highlight how Landau levels evolve from a Dirac cone when $B=0$ to completely flat bands when $l_B \ll \lambda$. This is analogous to keeping $B$ fixed and increasing the size of the flake, but the latter method is strongly constrained by numerical resources.
Here $l_B = 4.0$, $2.8$ and $2.0\ \textrm{nm}$ at $B=41$, $82$ and $164$\,T respectively, whereas $\lambda \sim 30$\,nm.

     \item \ \textbf{Effect of \emph{uniform} mass term on LL spectrum}

    \begin{figure}[b!]
\makebox[\textwidth]{\includegraphics[width=183mm]{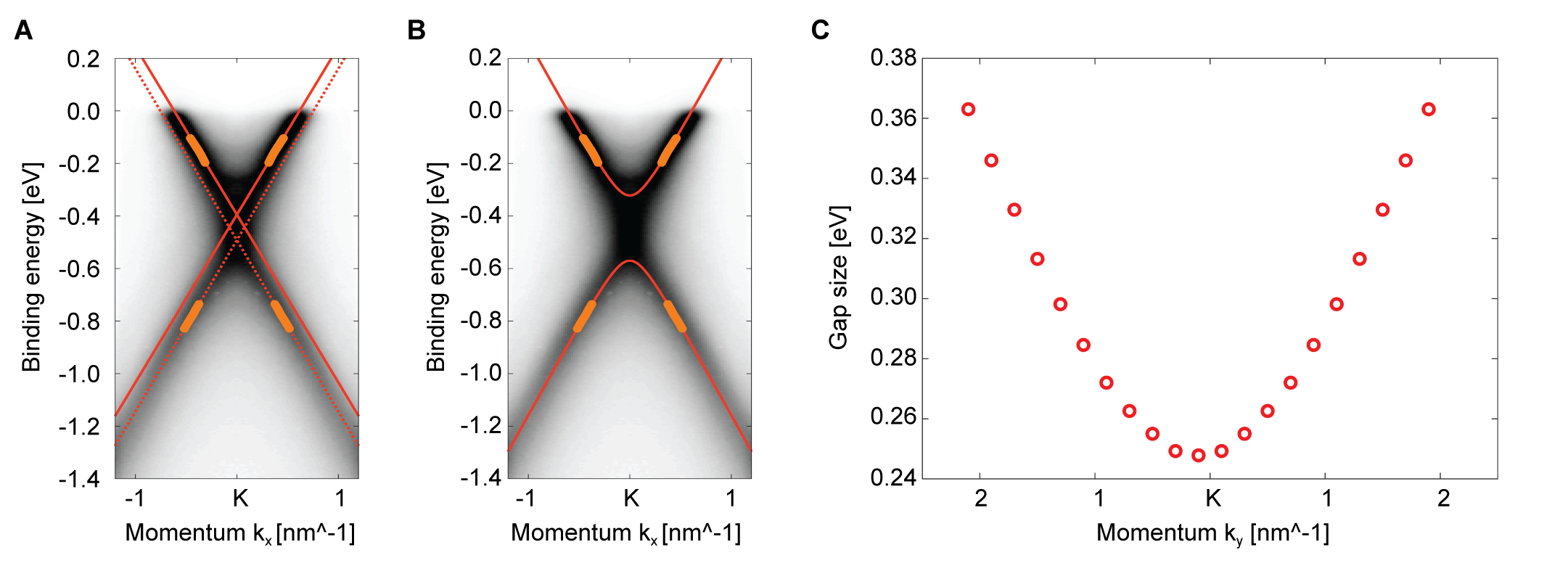}}
\caption{\textbf{Determination of the mass term.} (\textbf{A}) ARPES cut through the Dirac cone. Orange circles indicate the positions of the fitted Lorentzians. The red line and the dashed red line indicate linear fits through the orange circles for the upper and lower cone respectively. The cut is symmetrized around the K point in the momentum direction to remove polarization effects. (\textbf{B}) The same data as in (\textbf{A}), but fitted to a hyperbola instead. (\textbf{C}) Results for the gap size from the hyperbola fits for different ARPES slices along k$_y$. The curve shows the expected half-hyperbola and the gap size of $\sim$0.25\,eV is given by the minimum.}
\label{fig:s3_mass_term}
    \end{figure}

Here we briefly discuss the effect of a Semenoff mass\cite{semenoff_condensed-matter_1984} $M$ on pseudo-LLs and show that a \emph{uniform} Semenoff mass cannot explain the observed spectrum. Starting from the linear dispersing bands in the Dirac cone without any magnetic fields, a mass term opens a gap at the Dirac point. The size of the gap is equal to \emph{twice} the size of the mass term $M$. Experimentally, the existence of an inversion-breaking potential -- responsible for such a mass term -- has been proposed previously in the graphene on SiC sample system\cite{zhou_substrate-induced_2007}. It manifests in our ARPES cuts through the Dirac point by extending the linear dispersions of the lower and upper cones, for both sides with respect to the K point (Fig.\,\ref{fig:s3_mass_term}A), in that these extrapolations do not meet in a single point, but are offset from each other. To accurately determine the size of the gap, we fit two Lorentzians with a constant background to momentum distribution curves (MDCs) in the upper and lower cones. The energy range of the fit is selected to avoid the prominent LLs. A hyperbola is then fitted to the bands (Fig.\,\ref{fig:s3_mass_term}B) to determine top and bottom of the two bands, and in turn the gap size. The procedure is repeated for several cuts through the Dirac cone along the k$_y$ direction. The results are summarized in Fig.\,\ref{fig:s3_mass_term}C and the mass term is equal to half of the minimal gap size ($\sim$0.25\,eV). This is comparable to the $\sim$0.26\,eV gap observed in the same sample system by Zhou \emph{et al.}\protect\cite{zhou_substrate-induced_2007}.

    \begin{figure}[t!]
\makebox[\textwidth]{\includegraphics[width=89mm]{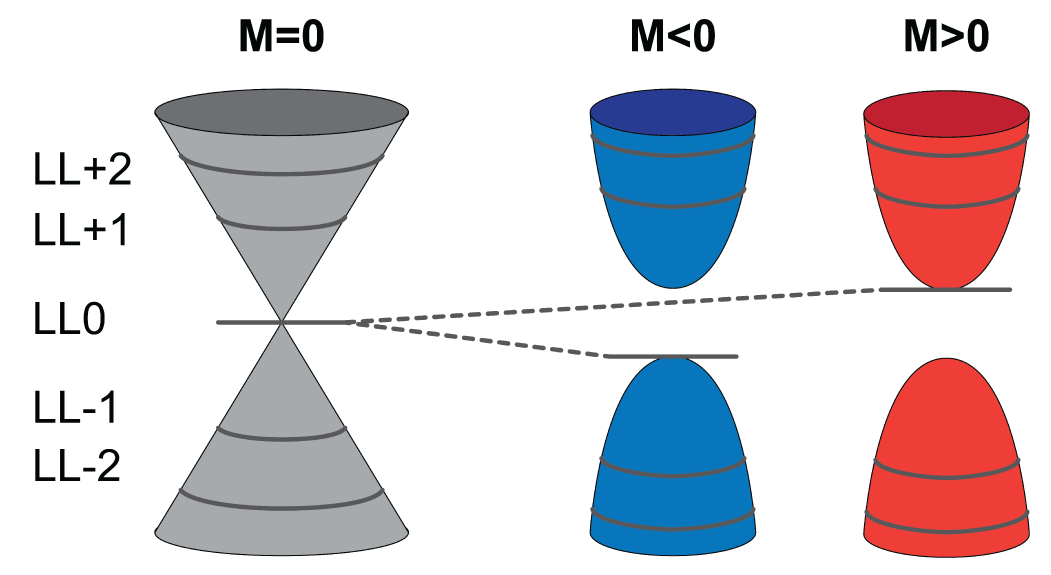}}
\caption{\textbf{Sketch of pseudo-LLs with Semenoff mass.} Depending on the sign of the mass term $M$, the zeroth LL (LL0) gets shifted to the upper or lower part of the cone. The spectrum is identical for valleys $K$ and $K'$, because pseudomagnetic fields preserve time-reversal symmetry. Higher LLs only get pushed away slightly from the Dirac point.}
\label{fig:s6_sketch_mass_term}
    \end{figure}

Next, we describe the effects of a mass term on a Dirac dispersion including magnetic fields. In this case the zeroth LL (LL0), which normally resides at the Dirac point, is gapped out and shifts by an energy equal to the mass term. Note that Eqn.\,\ref{eq:landaulevels_mass} is not properly defined for $n = 0$ -- to understand whether LL0 is shifted to $+M$ or $-M$ (in valleys $K$ and $K'$), we have to distinguish between real magnetic fields, which break time-reversal symmetry, and pseudomagnetic fields, which preserve time-reversal symmetry. For real magnetic fields\cite{hunt_massive_2013}, LL0 has opposite energy $\pm M$ at $K$ and $K'$. For pseudomagnetic fields, in order to preserve time-reversal symmetry, the spectrum must be identical in both valleys, and the energy of LL0 is determined by the sign of $M$, so for $n=0$ we simply get $E_{LL0}=E_{DP}\pm M$. This is illustrated in Fig.\,\ref{fig:s6_sketch_mass_term} for different signs of the mass term, where LL0 either shifts to the top of the lower cone ($M<0$) or the bottom of the upper cone ($M>0$).

     \begin{figure}[t!]
\makebox[\textwidth]{\includegraphics[width=183mm]{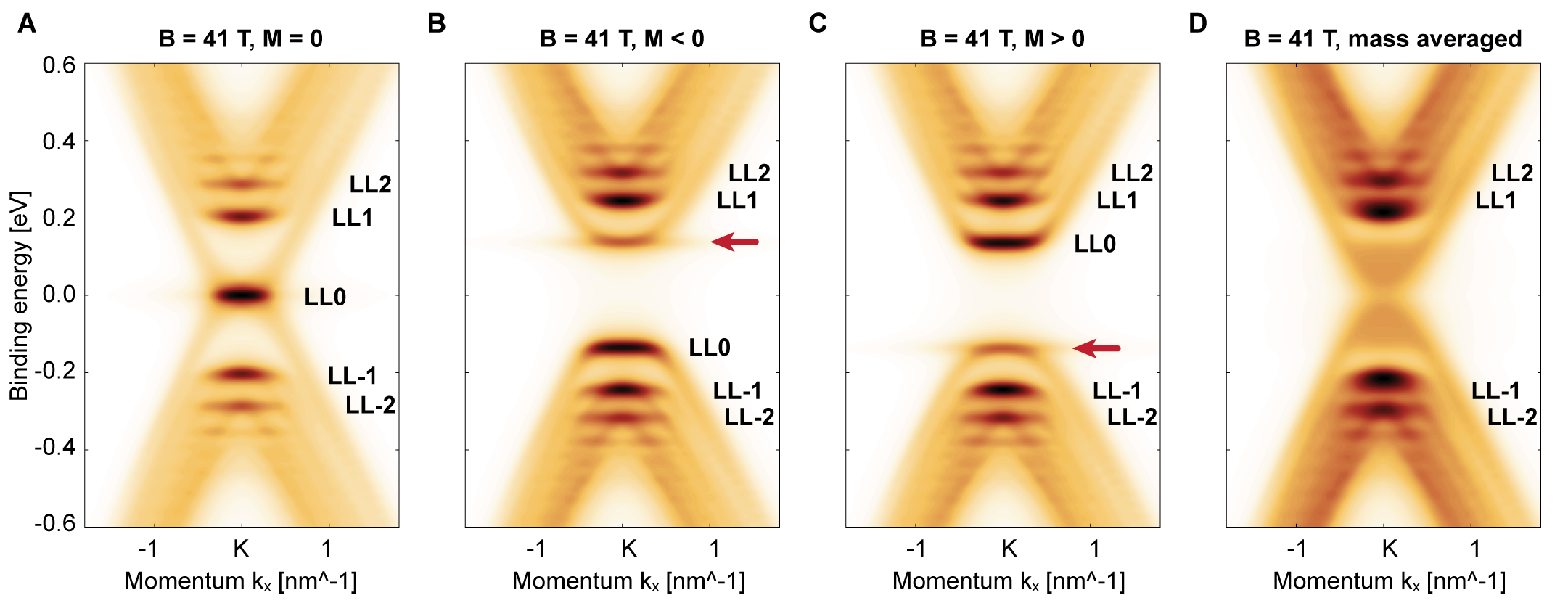}}
\caption{\textbf{Calculation of pseudo-LLs with Semenoff mass.} Calculated spectral function in our triangular flake with a uniform pseudomagnetic field $B=41$\,T and Semenoff masses $M = 0$, $M = -135$\,meV, $M = +135$\,meV, and averaged in the interval $M \in [-135, 135]$\,meV (from left to right). The position of the Landau levels (LL) for the different cases is indicated, as well as the much weaker LL0 from the area surrounding the strained flake (red arrows in (\textbf{B}) and (\textbf{C})).}
\label{fig:s10_model_mass_term}
    \end{figure}

Our numerical simulations clearly show this behaviour (Fig.\,\ref{fig:s10_model_mass_term}), but there is one additional caveat. The total pseudomagnetic flux must be vanishing in our flake by construction, as we require the strain to relax at the edges of the flake. This requirement generates a region near the boundaries of the strained area with a pseudomagnetic field of the reversed sign. This region hosts a LL0 at an energy inverted with respect to the LL0 coming from inside the strained area. This is visible in our calculations as weaker and more broadened (in momentum) levels, indicated by red arrows in Figs.\,\ref{fig:s10_model_mass_term}B and \ref{fig:s10_model_mass_term}C. Note that experimentally a similar scenario is natural on our graphene on SiC samples as well. The strain inside the nanoprisms needs to relax away from the feature, thus creating an area with an inversed pseudomagnetic field. 
\linebreak To check if a \emph{uniform} mass term of about the determined size can explain our findings, we fit the observed LLs to Eqn.\,\ref{eq:landaulevels_mass} (see Fig.\,\ref{fig:s4_fit_with_mass}). While this model produces a qualitatively good fit with $M=150\,\textrm{meV}$, it places the Dirac point at a binding energy of $390\,\textrm{meV}$, which is inconsistent with the experimental observations (compared to $450\,\textrm{meV}$ obtained from the fit to Eqn.\,\ref{eq:landaulevels} without a mass term). Hence, in order to explain the absence of a sharp LL0 in the ARPES data, we instead postulate that the mass term $M$ varies slowly with respect to the magnetic length $l_B$, as discussed in the main text. This variation can take place either from nanoprism to nanoprism, or within a given nanoprism, if it is tied to the length scale of the uniform pseudomagnetic field $\lambda$. In this scenario, we can approximate the effect of the slowly-varying mass term $M$ by averaging over the spectral function obtained with different fixed $M$ (such as those shown in Figs.\,\ref{fig:s10_model_mass_term}B and \ref{fig:s10_model_mass_term}C). This mechanism completely smears out LL0, while only slightly broadening the other levels (see Fig.\,\ref{fig:s10_model_mass_term}D).

    \begin{figure}[t!]
\makebox[\textwidth]{\includegraphics[width=89mm]{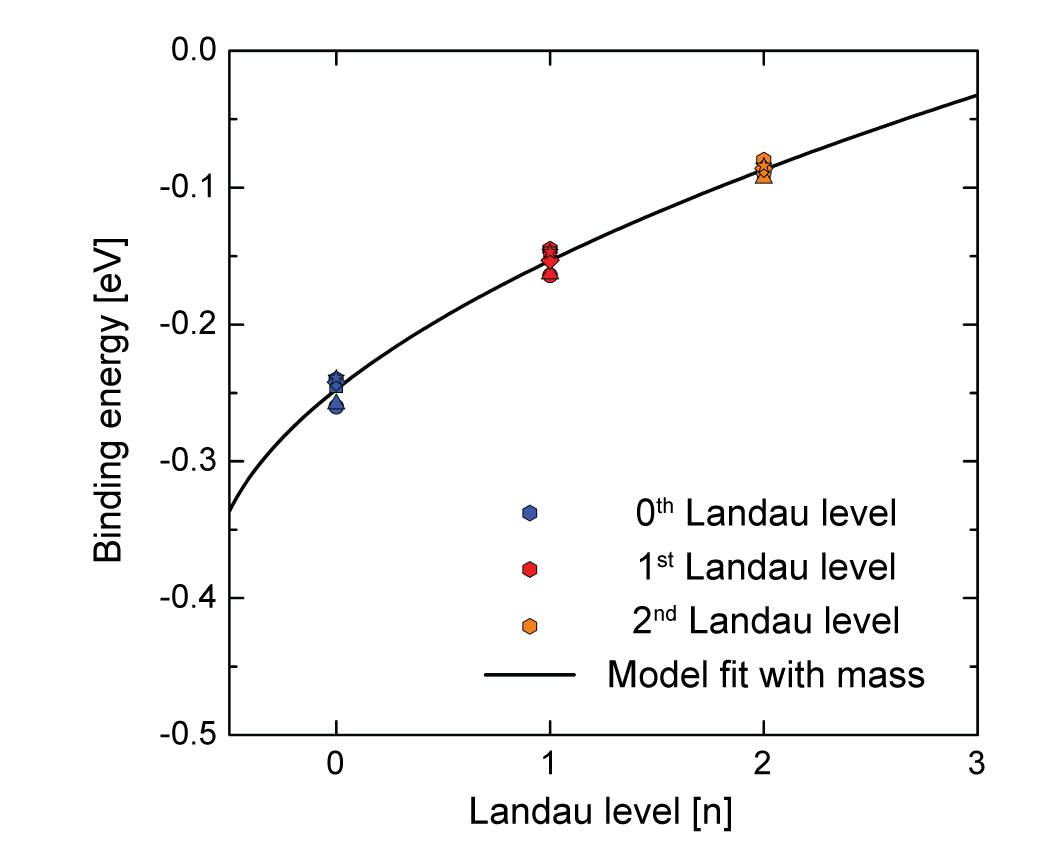}}
\caption{\textbf{Model fit with constant mass term.} Fit of the observed LLs to Eqn.\,\ref{eq:landaulevels_mass}. Note the shifted indices for the LLs in this scenario. It places the Dirac point at a binding energy of 390\,meV with $M=150\,\textrm{meV}$, compared to 450\,meV obtained from the fit to Eqn.\,\ref{eq:landaulevels} without a mass term.}
\label{fig:s4_fit_with_mass}
    \end{figure}

\pagebreak

    \item \ \textbf{Additional ARPES data}
    
    \begin{figure}[b!]
\makebox[\textwidth]{\includegraphics[width=120mm]{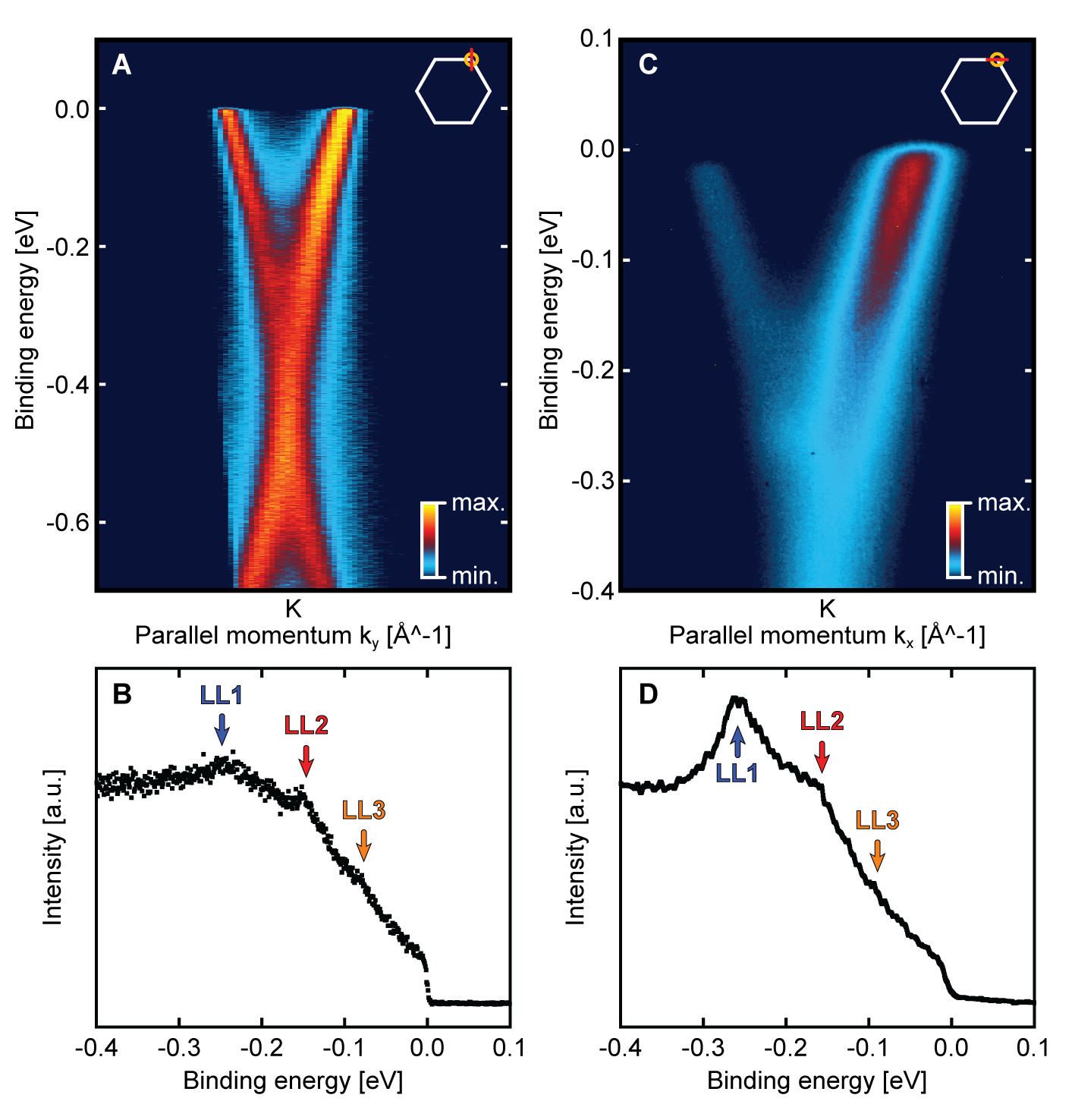}}
\caption{\textbf{ARPES data on two additional samples.} (\textbf{A}) ARPES cut along the direction indicated by the red line in the schematic BZ in the top right corner. Data were taken at 6\,K with mostly unpolarized light. (\textbf{B}) Energy cut through the Dirac point of the data in (\textbf{A}) with LLs marked with arrows. (\textbf{C}) ARPES cut along the direction indicated by the red line in the schematic BZ in the top right corner. Data were taken at 6\,K with p-polarized light. (\textbf{D}) Energy cut through the Dirac point of the data in (\textbf{C}) with LLs marked with arrows. Both data sets are unsymmetrized.}
\label{fig:r1_additional_arpes}
    \end{figure}
    
ARPES data for two additional samples complementary to the data in Fig.\,\ref{fig:f1_arpes} is shown in Fig.\,\ref{fig:r1_additional_arpes} with LLs indicated in the cuts along the energy axis. The data have not been symmetrized and the LLs are still clearly visible in the energy cuts. While the APRES data in the main text (Fig.\,\ref{fig:f1_arpes}A) were acquired with s-polarized light, the data in Fig.\,\ref{fig:r1_additional_arpes}A were taken with mostly unpolarized light, and the data in Fig.\,\ref{fig:r1_additional_arpes}C were taken with p-polarized light. The different light polarizations change the ARPES intensity distribution due to matrix element effects, but do not alter the position of the observed LLs in the energy cuts. For unpolarized light an almost symmetric intensity distribution for both branches of the Dirac cone can be observed, even without additional symmetrization (see Fig.\,\ref{fig:r1_additional_arpes}A).

     \item \ \textbf{Nanoprism distribution and step edge}
     
     \begin{figure}[b!]
\makebox[\textwidth]{\includegraphics[width=89mm]{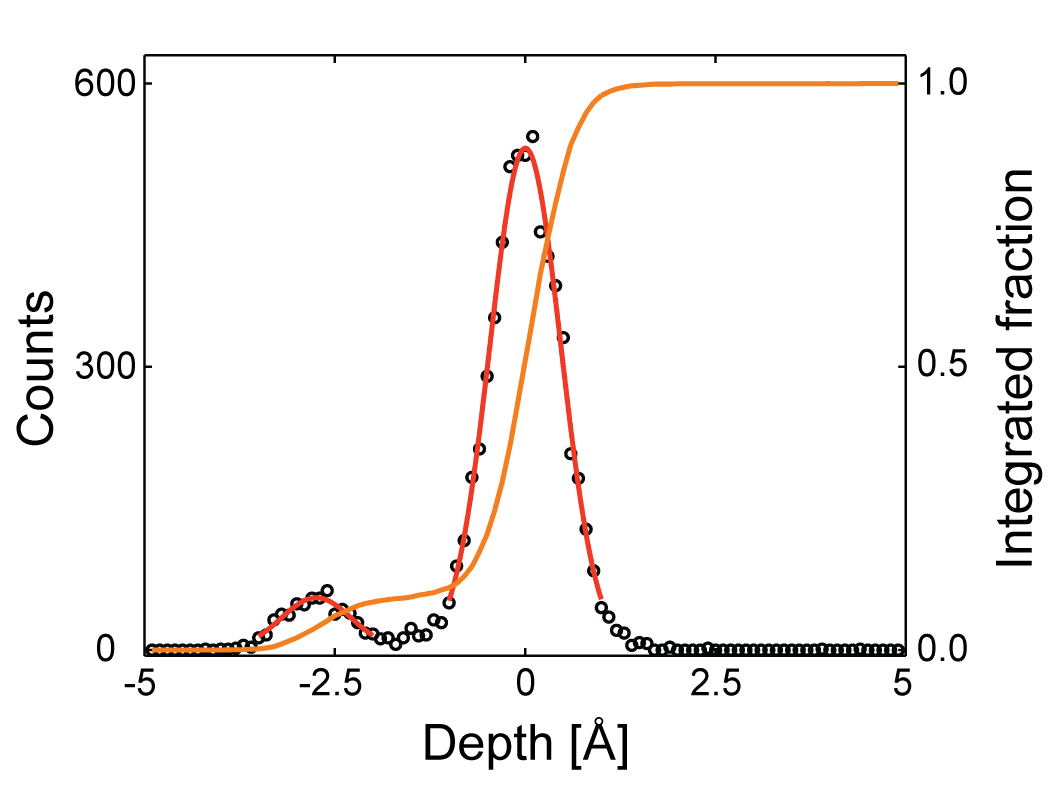}}
\caption{\textbf{AFM height distribution.} Height distribution for the AFM image in Fig.\,\ref{fig:f2_afm}B (Top). Two Gaussians (red) can be fitted to the data to extract the depths of the nanoprisms. The integrated fraction curve is shown in yellow.}
\label{fig:s9_height_distribution}
    \end{figure}

Looking at the height distribution of the pixels in the AFM image in Fig.\,\ref{fig:f2_afm}B (Top), we can determine the depths of the nanoprisms as wells as estimate the coverage of the nanoprisms on the sample (Fig.\,\ref{fig:s9_height_distribution}). The difference in the position of the two fitted Gaussians leads to a depth of the nanoprisms of $(2.7\pm 0.7)\,\textrm{\AA}$. The integrated fraction curve indicates that about 5\% to 10\% of the total area is covered with nanoprisms. 

The STM image taken across the edge of a nanoprism in Fig.\,\ref{fig:s13_triangle edge}A shows how the graphene grows smoothly over the step without interruption. This assures that the strain inside the nanoprism can build up and is not relieved along grain boundaries. Adhesion measurements (see Methods section) unambiguously distinguish between coverages of zero-, mono-, and bilayer graphene\protect\cite{emtsev_interaction_2008,virojanadara_homogeneous_2008}. The AFM adhesion image in Fig.\,\ref{fig:s13_triangle edge}B (taken in the same region as in Fig.\,1A) shows no contrast between the nanoprisms and the surrounding terraces, thus clearly indicating that the nanoprisms are covered by monolayer graphene.

        \begin{figure}[t!]
\makebox[\textwidth]{\includegraphics[width=183mm]{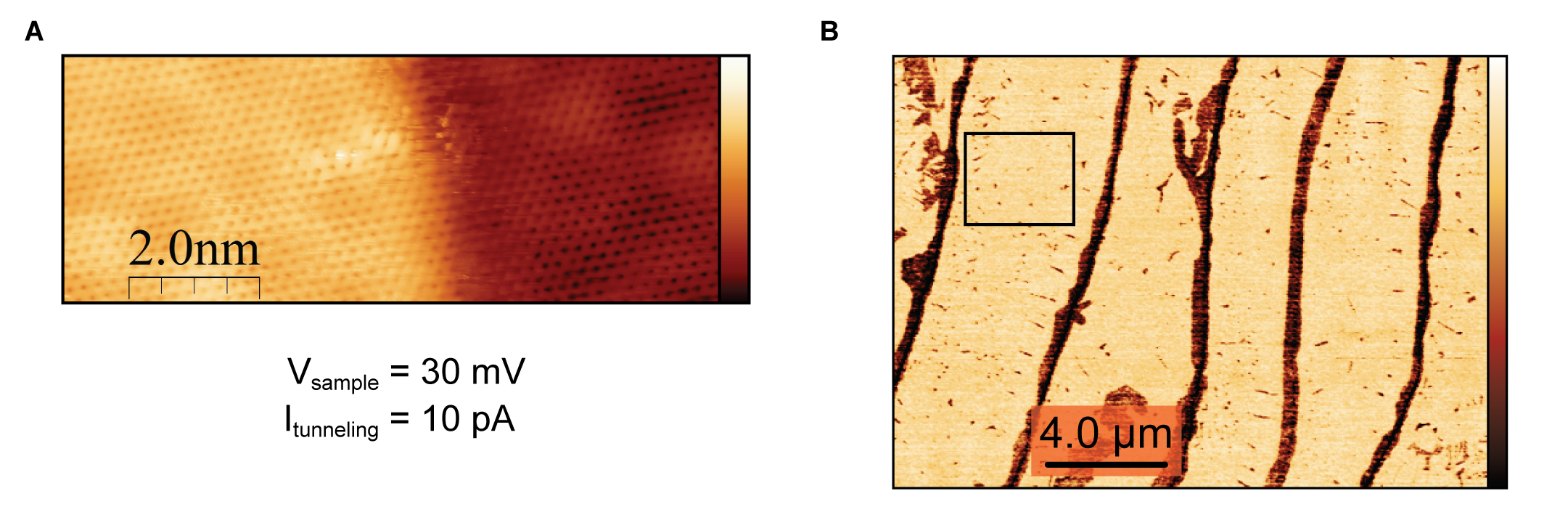}}
\caption{\textbf{Graphene layer coverage.} (\textbf{A}) STM image taken across the edge of a nanoprism ($V_{sample}=30\,\textrm{mV}$, $I_{tun.}=10\,\textrm{pA}$). The graphene grows smoothly over the step without interruption. (\textbf{B}) AFM adhesion image taken in the same region as shown in Fig.\,\ref{fig:f2_afm}A in the main text. The image shows no contrast between the nanoprisms and the surrounding terraces (black box), thus clearly indicating that the nanoprisms are covered by monolayer graphene.}
\label{fig:s13_triangle edge}
    \end{figure}

\end{enumerate}

\end{document}